\documentclass[sigconf]{acmart}
\usepackage[T2A]{fontenc}
\usepackage[utf8]{inputenc}

\AtBeginDocument{%
  \providecommand\BibTeX{{%
    \normalfont B\kern-0.5em{\scshape i\kern-0.25em b}\kern-0.8em\TeX}}}

\copyrightyear{2021}
\acmYear{2021}
\setcopyright{iw3c2w3}
\acmConference[WWW '21 Companion]{Companion Proceedings of the Web Conference 2021}{April 19--23, 2021}{Ljubljana, Slovenia}
\acmBooktitle{Companion Proceedings of the Web Conference 2021 (WWW '21 Companion), April 19--23, 2021, Ljubljana, Slovenia}
\acmPrice{}
\acmDOI{10.1145/3442442.3452306}
\acmISBN{978-1-4503-8313-4/21/04}

\begin{document}

\title{Auditing Source Diversity Bias in Video Search Results Using Virtual Agents}

\author{Aleksandra Urman}
\email{aleksandra.urman@ikmb.unibe.ch}
\affiliation{%
  \institution{University of Bern}
  \institution{University of Zurich}
  \country{Switzerland}
}

\author{Mykola Makhortykh}
\email{mykola.makhortykh@ikmb.unibe.ch}
\affiliation{%
  \institution{University of Bern}
  \country{Switzerland}
}

\author{Roberto Ulloa}
\email{roberto.ulloa@gesis.org}
\affiliation{%
  \institution{GESIS – Leibniz Institute for the Social Sciences}
  \country{Germany}
}

\renewcommand{\shortauthors}{Urman et al.}

\begin{abstract}
    We audit the presence of domain-level source diversity bias in video search results. Using a virtual agent-based approach, we compare outputs of four Western and one non-Western search engines for English and Russian queries. Our findings highlight that source diversity varies substantially depending on the language with English queries returning more diverse outputs. We also find disproportionately high presence of a single platform, YouTube, in top search outputs for all Western search engines except Google. At the same time, we observe that Youtube's major competitors such as Vimeo or Dailymotion do not appear in the sampled Google's video search results. This finding suggests that Google might be downgrading the results from the  main competitors of Google-owned Youtube and highlights the necessity for further studies focusing on the presence of own-content bias in Google's search results.
\end{abstract}



\maketitle

\section{Introduction}

The public tends to put high trust in the information retrieved via search engines \cite{noauthor_2020_nodate}. However, search engine outputs are prone to different forms of bias \cite{kay_unequal_2015,goldman_search_2008} which can result in distorted representation of subjects which users are searching for \cite{grimmelmann_skepticism_2010, mowshowitz_assessing_2002}. One way of uncovering bias in web search outputs is to engage in
algorithmic auditing \cite{mittelstadt_automation_2016}. Though a few studies have audited bias in text and image search (e.g., \cite{kay_unequal_2015,otterbacher_competent_2017,kulshrestha_search_2019}), to the best of our knowledge no audits of video search results were conducted. 
However, video search is important in the societal context since people increasingly consume news via online videos \cite{kalogeropoulos_online_2018} and treat video hosting platforms as a preferred environment for news finding \cite{stocking_many_2020}. The fact that video information can have powerful influence on users and even affect their behaviors \cite{izawa_what_nodate,erickson_viral_2018} prompts the need for auditing whether video search is subjected to bias.

To address this gap, we investigate the presence of source diversity bias - that is a systematic prevalence of specific information sources \cite{goldman_search_2008} - in video search outputs. By consistently prioritizing the same set of sources independently of search queries, search algorithms can diminish the quality of outputs by making them less representative \cite{mowshowitz_assessing_2002} and negatively affecting the user experience \cite{ciampaglia_how_2018}. Source diversity bias is also related to the phenomenon of search concentration, namely the tendency of search engines to prioritize few well-established domains over other sources \cite{jiang_search_2014} that, according to media, often results in search companies promoting their own services (e.g., YouTube in the case of Google \cite{west_searching_2020}). By diminishing the diversity in the composition of source domains, companies can consolidate global media monopolies \cite{jiang_search_2014} through gaining unfair advantage over their competitors.

To examine whether video search outputs are subjected to source diversity bias, we audit search results coming from five search engines - four Western (Bing, DuckDuckGo, Google, and Yahoo) and one non-Western (Yandex) - in response to English and Russian queries. Including Yandex along with queries in Russian - a language dominating the main markets of Yandex - allowed us to test whether (some of) our observations can be attributed to structural differences between Western and non-Western markets (e.g., almost monopolistic status of Google in the former and its competition with Yandex in post-Soviet countries). We use virtual agent-based auditing approach to prevent search outputs from being affected by search personalization and search randomization. Then, using a selection of metrics, we assess the level of source domain diversity in search outputs and investigate whether there is evidence of certain engines prioritizing specific information sources. Specifically, we examine whether search engines tend to promote platforms associated with their parent companies (e.g., Alphabet for Google or Microsoft for Bing) or downgrade the competitors as was claimed by earlier research \cite{jiang_search_2014}.


\section{Related work}

The problem of auditing systematic skewness of web search outputs is increasingly recognized in the field of information retrieval (IR) \cite{mowshowitz_assessing_2002,grimmelmann_skepticism_2010,goldman_search_2008,introna_shaping_2000}. Existing studies primarily look at it from one of the two perspectives: user bias and retrieval bias. User bias concerns skews in user perceptions of search outputs \cite{ieong_domain_2012}. Retrieval bias relates to a skewed selection of search results \cite{introna_shaping_2000,grimmelmann_skepticism_2010}.

One form of retrieval bias, which the current paper focuses on, is source diversity bias. Originally discussed in the context of search engines' tendency to prioritize web pages with the highest number of visitors \cite{introna_shaping_2000}, source diversity bias is currently investigated in the context of prioritization of certain categories of sources in response to particular types of queries (e.g., \cite{makhortykh_how_2020}). A disproportionate visibility of specific types of web resources can diminish overall quality of search results \cite{ciampaglia_how_2018} and provide unfair advantage to companies and individuals that own specific search engines \cite{introna_shaping_2000} - e.g., through own-content bias, - or direct most of the traffic to a handful of well-established sources, a phenomenon also known as search concentration \cite{jiang_search_2014}.

To date, source diversity bias has been primarily investigated in the context of text search results \cite{diakopoulos_i_nodate,unkel_googling_2019,trielli_search_2019} with the focus exclusively on one engine - Google. At the same time, few comparative studies that were conducted highlight substantial cross-engine differences in search diversity bias levels \cite{jiang_business_2014,zavadski_querying_2019,makhortykh_how_2020}. In the case of video search results, there is no systematic comparative assessment of source diversity bias. In 2020, the Wall Street Journal \cite{west_searching_2020} and, subsequently, Mozilla \cite{noauthor_youtube_nodate} have found that Youtube appears among the top-3 featured "carousel" results in text search 94\% of the time \cite{noauthor_youtube_nodate}. These findings highlight search concentration \cite{jiang_search_2014} around Youtube in Google's "carousel" results. However, it is unclear, first, whether the distribution of domains is similar in dedicated video search results. And second, how Google's results compare to those of its competitors - a comparison that is necessary to establish whether Youtube's dominance in video search results from the way Google's algorithm works exhibiting own-content bias. We aim to address these gaps with the present study.

\section{Methods}
In this study, we have opted for a combination of methods for bias detection in web search outlined by Edelman \cite{edelman_bias_nodate}: 1) comparative analysis of the results provided by multiple search engines across a variety of search queries in two languages; 2) identification of skewness of results towards specific domains and whether the skewness, if observed, can be explained by market structure incentives. We have used video search results from the 4 biggest Western search engines by market share - Google, Yahoo, Bing, DuckDuckGo, - and one major non-Western search engine - Yandex \cite{noauthor_search_nodate}. Since Yandex has the largest presence in post-Soviet states where large shares of populations speak Russian as a first language, we utilized queries in both, English and Russian, to conduct the search and estimate whether there are differences in the observations. There were 62 queries in total - 31 English queries with translations into Russian, - that concerned contemporary events (i.e., the US presidential elections and coronavirus), conspiracy theories (i.e., Flat Earth), and historical events (i.e., Holocaust)\footnote{We used the following queries: \textit{coronavirus, bernie sanders, joe biden, pete buttigieg, elizabeth warren, michael bloomberg, donald trump, us elections, syria conflict, ukraine conflict, yemen conflict, holocaust, holodomor, armenian genocide, second world war, first world war, artificial intelligence, big data, virtual reality, vaccination, vaccination benefits, vaccination dangers, george soros, illuminati, new world order, Flat Earth, UFO, Aliens, misinformation, disinformation, fake news}. All queries were entered into search engines in lower case. For the searches in Russian, we used the exact translations of the English queries listed below into Russian verified by two native Russian speakers. }. Below we outline the details on the data collection and analysis.

\subsection{Data collection}
To collect the data, we utilized a set of virtual agents - that is software simulating user browsing behavior and recording its outputs. The benefits of this approach, which extends algorithmic auditing methodology introduced by Haim et al. \cite{haim_abyss_2017}, is that it allows controlling for personalization \cite{hannak_measuring_2013} and randomization \cite{makhortykh_how_2020} factors. 


For the current study, we built a network of 84 CentOS virtual machines based in the Frankfurt region of Amazon Elastic Compute Cloud (EC2). On each machine, we deployed 2 virtual agents (one in Chrome browser and one in Mozilla Firefox browser), thus providing us with 188 agents overall. Each agent was made of two browser extensions: a tracker and a bot. The tracker collected the HTML and the metadata of all pages visited in the browser and immediately sent it to a storage server. The bot emulated a sequence of browsing actions that consisted of (1) visiting a video search engine page, (2) entering one of the 62 queries, and (3) scrolling down the search result page to load at least 50 images. Before searching for a new query, the browsers were cleaned to prevent the search history affecting the search outputs, and there was a 7-minute break between searches to mitigate potential effects of previous searches on the results.

The study was conducted on February 26, 2020. We equally distributed the agents between five search engines: Google, Bing, Yahoo, Yandex, and DuckDuckGo (DDG)\footnote{For all engines, the ".com" version of the image search engine was used (e.g., google.com).}. Because of technical issues (e.g., bot detection mechanisms), some agents did not manage to complete their routine. The overall number of agents per engine which completed the full simulation routine and returned the search results differed by query - sometimes the search engine would detect automation and temporarily ban the agent. This was particularly often the case with Yandex, where for some queries all 34 deployed agents successfully finished the routine while for others (a minority of queries) Yandex only returned the results for 10 agents and banned the rest. The mean number of agents who completed the full routine by engine across all queries is the following: Bing (29), DDG (34), Google (33), Yahoo (31), and Yandex (17). 

After the data was collected, we have extracted top-10 individual video links obtained by each agent for each search query and proceeded with the analysis using this data. Our decision to rely on the top-10 results only is motivated by the fact that users tend to pay the most attention to the first few results - i.e., those on the first results page \cite{pan_google_2007}. A comparison of search results by browser has demonstrated that there are no major between-browser differences - a finding in contrast with those observed for text search results \cite{makhortykh_how_2020} - thus for the analysis we have proceeded aggregating the results for both browsers.

\subsection{Data analysis}
\subsubsection{Source diversity}
To assess whether there is evidence suggesting that diversity bias - that is, lack of source diversity, - is present in the sampled results on domain level, we have calculated how many distinct source domains are, on average, present in the results for each query. To account for the potential randomization due to so-called "Google Dance" \cite{battelle_search_2005} in the results, for each query we aggregated the calculation over the results obtained from each individual autonomous agent. Afterwards, we have calculated mean numbers of distinct sources separately for English and Russian queries to establish whether there are differences in the observations depending on the language of the search. We have also qualitatively examined the results per query to find out whether there are distinct patterns with regard to query categories.

\subsubsection{Search concentration}
To establish whether there is evidence of search concentration in the collected video results, we have calculated, first, the share of times different domains appear as the top result for each search query, and second, the proportion of times different domains appear among the top-10 search results per query at all. We suggest that the consistent appearance of a specific domain or few specific domains at the very top of search results and, in general, among top-10 results more frequently than the others would indicate search concentration.

In addition, we have scrutinized the results with regard to own-content bias exhibited by Google according to the media \cite{west_searching_2020}. We aimed to establish whether Google's results lend evidence of own-content bias either through the promotion of Youtube in the results or the demotion of the results provided by its main competitors. One way to assess that is to compare Google's results with those obtained through other search engines \cite{edelman_bias_nodate} and see whether there are major differences in the observed frequencies of appearance of different domains between them. Thus, we compared the proportion of times each domain - Youtube and its competitors such as Vimeo, Dailymotion and Rutube, - appears in the results provided by different engines.

\section{Findings}
\subsection{Source diversity}

\begin{figure}[h]
  \centering
  \includegraphics[width=\linewidth]{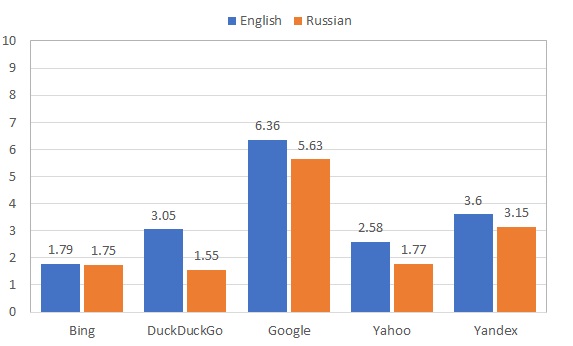}
  \caption{Mean number of distinct domains in top-10 video search results (Y-axis) per query, grouped by query languages (legend) and search engines (X-axis)}
  \label{fig0}
\end{figure}

As Fig.\ref{fig0} shows, there are differences in the level of source diversity exhibited by the examined search engines. Google has consistently presented more diverse video results in terms of source domains than its competitors. Yandex has taken a second place in terms of domain diversity. This domain diversity hierarchy is similar for both English and Russian queries, albeit for the Russian queries the observed domain diversity is a bit lower on all five search engines than for the English queries.

Qualitative analysis of the domain diversity by query has demonstrated that there are no consistent patterns with regard to the proportion of distinct sources by query category in our dataset. Hence, we suggest that domain diversity is affected more by the algorithms used by each search engine examined and, probably, the data they are trained on - the latter might explain the observed differences between Russian and English queries, - rather than by the specific topics of the search queries.

\subsection{Search concentration}
\begin{figure}[h]
  \centering
  \includegraphics[width=\linewidth]{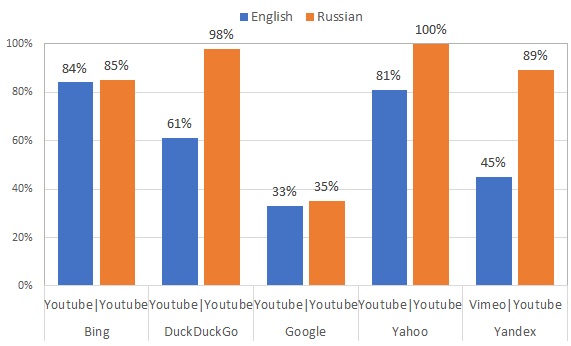}
  \caption{Domain most frequently appearing as the top result by query group and search engine; \% of time it appears as the top result.}
  \label{fig1}
\end{figure}

As shown in Fig.\ref{fig1}, Youtube has been featured as the top result most frequently in all cases but one - namely, the results for English queries on Yandex where Vimeo surfaced as the first result most often. Remarkably, on Google itself Youtube appeared as the top result less frequently than on other platforms, a finding in contrast with those made by The Wall Street Journal \cite{west_searching_2020} and Mozilla \cite{noauthor_youtube_nodate} in the context of featured video "carousel" on the first page of Google's text results. DuckDuckGo, Yahoo, Yandex are the three engines exhibiting sizeable differences in the prominence of Youtube as the top result between English and Russian queries with Youtube being featured as the top result more frequently in response to English queries. We suggest that the findings reported in Fig.\ref{fig1} lend evidence to search concentration bias in video search results on the examined search engines, with the effect in our sample being stronger for English than for Russian queries.

\begin{figure*}
    \centering
    \resizebox{\textwidth}{!}{\includegraphics{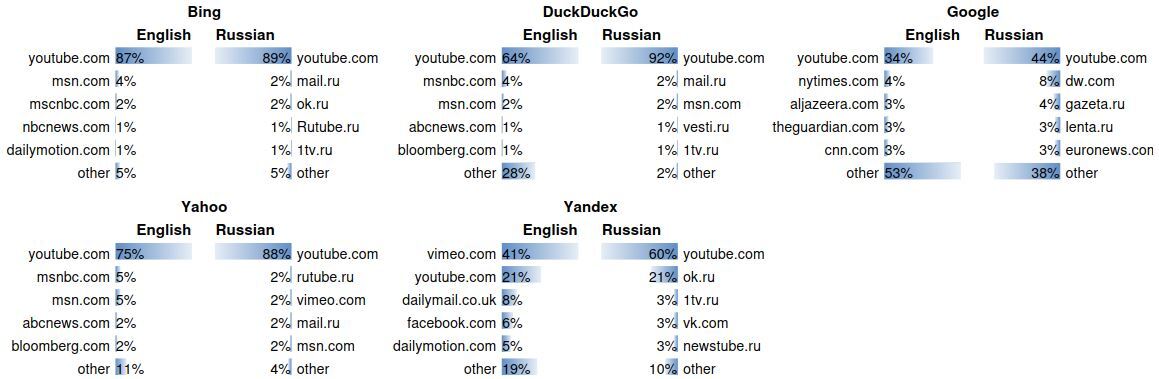}}
    \caption{Domains most frequently appearing among top-10 results by query group and search engine and \% of time a domain appears among top-10 results.}
    \label{fig2}
\end{figure*}

In Fig.\ref{fig2} we list the domains most frequently featured among top-10 search results on each of the examined search engines for English and Russian queries. As with the top-1 result, the domain most frequently featured in top-10 is Youtube on all engines but Yandex in response to English queries where the most frequent domain is Vimeo. The share of other domains in search results is comparatively marginal - less than 10\% - in almost all cases with the exception of Ok.ru for Russian queries on Yandex. Youtube thus emerges as the most prominent domain in search results. Apart from it, there is no other domain that is among the top 5 domains most frequently appearing among the first 10 results on all search engines. 

The findings reported in Fig.\ref{fig2} suggest that search engines feature different arrays of domains - with the exception of Youtube - in search results. Google tends to retrieve results from legacy media in both Russian and English more frequently than other search engines, a finding in line with the previous research \cite{trielli_search_2019,diakopoulos_i_nodate,unkel_googling_2019}. Other search engines also include some legacy media, though to a lower extent than Google, as well as social media (e.g., Ok.ru, Facebook.com), and several video portals that are Youtube's competitors - Dailymotion, Vimeo and Rutube. None of these potential Youtube competitors, however, appear in the top-10 results on Google in our dataset at all despite their presence on other search engines. Vimeo, Dailymotion and Rutube all appear at least once among the top-10 results on all search engines except Google. This finding suggests that Google might downgrade Youtube's direct competitors, however an analysis based on a broader spectrum of queries is necessary to estimate the scope and persistence of this result.

\section{Conclusions and future work}

Top 10 outputs of video search for most of search engines except Google show limited source diversity. By relying on average on 2-3 unique sources to retrieve top results for English queries, search engines create a situation in which users' information choices are shaped by a few content providers. This raises concerns about search engines facilitating consolidation of power on the information markets.

The only exception among the five search engines examined is Google, where the degree of source diversity is almost twice as high. This effect can be attributed to Google putting substantial effort into diversifying search results in response to earlier criticisms. With 6 unique domains per 10 top results, Google follows its declared principle of having no more than 2 results coming from the same domain in the top results \cite{noauthor_google_2019}. The finding also suggests that low diversity on other search engines is likely attributed to the absence of diversification mechanisms which Google implements.

The importance of integrating diversification measurements is highlighted by high degree of source diversity bias. The top results for all search engines (except Yandex in English and Google) are dominated by one platform, namely YouTube. Its systematic prevalence reinforces the platform's almost monopolistic status. It is problematic considering that the platform is already used as a major news source among certain shares of the population \cite{kalogeropoulos_online_2018}, despite earlier audits demonstrating that its algorithms might lead to user radicalization \cite{ribeiro_auditing_2020} and, affected by users' viewing history, aggressively promote pseudoscientific content to users who have watched pseudoscientific videos before \cite{papadamou_it_2020}. Search concentration around Youtube can only help cement its monopoly and the associated effects.


Google fairs better than other search engines in terms of domain-level source diversity and, at a first glance, does not exhibit own-content bias since Youtube is less prominent in its results than on other search engines. However, in our sample, Google is the only search engine that did not provide any results from Youtube's major competitors. It is thus plausible that Google indeed, as media reports suggested \cite{west_searching_2020}, might be downgrading the results coming from the major competitors of Youtube thus exhibiting own-content bias manifested not in promoting Youtube but in lowering the prominence of its competitors in search results. However, it could also be that the obtained results with regards to Google and the absence of Youtube's competitors in its outputs are specific to the topics addressed by our sample of queries and is absent in other contexts. Hence, to make a robust conclusion about the presence or absence of own-content bias in Google's video search results, further studies encompassing broader sets of queries are necessary. We suggest that our observations highlight the need for such studies.

Further, the present analysis is based on a snapshot experiment on a limited selection of queries. We believe that our findings warrant subsequent longitudinal audits of video search results to assess the persistence of our observations and, potentially, the changes that occur overtime. Such audits are crucial to inform the decisions of policy-makers and regulators. This is especially timely and pressing given the recent anti-trust cases against Google \cite{edelman_googles_nodate} and calls for putting tech giants under scrutiny, among other, in the context of their market power.

\bibliographystyle{ACM-Reference-Format}
\bibliography{FATES}




\end{document}